\documentclass[10pt,prl,aps,twocolumn,tightenlines,superscriptaddress,nofootinbib,notitlepage,preprintnumbers]{revtex4-1}
\pdfoutput=1
\usepackage{epsfig,amsfonts,mathrsfs,amsmath,amssymb,graphicx,color}

\begin{document}
\title{Hidden SU(N) Glueball Dark Matter}
\author{Amarjit Soni}
\affiliation{Physics Department, Brookhaven National Laboratory, Upton, NY 11973, USA}
\author{Yue Zhang}
\affiliation{Walter Burke Institute for Theoretical Physics,\\
California Institute of Technology, Pasadena, CA 91125, USA}

\begin{abstract}

We investigate the possibility that the dark matter candidate is from a pure non-abelian gauge theory of the hidden sector, motivated in large part by its elegance and simplicity. The dark matter is the lightest bound state made of the confined  gauge fields, the hidden glueball. We point out this simple setup is capable of providing rich and novel phenomena in the dark sector, especially in the parameter space of large $N$. They include self-interacting and warm dark matter scenarios, Bose-Einstein condensation leading to massive dark stars possibly millions of times heavier than our sun giving rise to gravitational lensing effects, and indirect detections through higher dimensional operators as well as interesting collider signatures.
\end{abstract}

\preprint{CALT-TH-2016-002}

\maketitle

\medskip
\noindent{\bfseries Introduction.}
An outstanding issue of fundamental importance in particle physics is the nature of the dark matter (DM).
This question is particularly intriguing and perplexing, given the preponderance of DM over visible matter and its profound gravitational effects throughout the evolution of the universe.

In this work, we like to investigate the viability of the dark matter candidate from the hidden sector with a non-abelian gauge symmetry, 
a minimal theory with non-trivial mass scale. 
The gauge group is chosen to be $SU(N)$, and, for simplicity, neither fermions nor any other particle is introduced
in that sector. The dark matter is the lightest hidden glueball state, which is likely a scalar field, and a non-perturbative bound state made of a pair of confined  gauge fields.
This is a very simple setup with only a handful of parameters, which are the intrinsic scale $\Lambda$, the number of colors $N$, and $\theta$\,---\,for the T and P-odd $\theta$-term in the hidden sector. They control the mass and all the couplings of the hidden glueball dark matter
(GDM), named $\phi$ hereafter. 

In spite of the simplicity of this setup, we will show that the hidden glueball indeed satisfies all the conditions for a dark matter candidate. Moreover, such a dark matter could be both self-interacting and warm, thus safely 
evading all the potential problems of the usual collisionless cold dark matter.
The scalar GDM could have the novel feature of Bose-Einstein condensation into compact objects thus 
plausibly leading to 
interesting gravitational effects such as microlensing. It could also be tested in particle physics experiments 
if there exist interactions of it with standard model particles
via higher dimensional operators.
We will elaborate on these points in order in the following sections~\cite{otherworks}.

\smallskip
\noindent{\bfseries Hidden Glueball as Dark Matter.}
In this work, we consider dark matter candidate (DMC) from  a very simple setup, a hidden sector non-abelian gauge symmetry with only gauge fields and without fermions. 
The Lagrangian of the model is
\begin{eqnarray}
\mathcal{L} = - \frac{1}{4} H^{a}_{\,\mu\nu} H^{a\,\mu\nu} \ . 
\end{eqnarray}
where $H^a_{\,\mu\nu}$ is the gauge field strength of the group $SU(N)$, with an unspecified value of $N$ to be determined later.
As is well known the gauge coupling $g_h$ becomes large 
at low energy scale and dimensional transmutation generates a scale $\Lambda$ for the theory, similar to the emergence of the QCD scale. Around the scale $\Lambda$, the physical degrees of freedom turn into a tower of hidden glueballs. 
From the knowledge based on existing 
calculations, the lowest lying glueball states when $\theta=0$ carry quantum numbers $J^{PC} = 0^{++}$, or $0^{-+}$~\cite{Cornwall:1982zn, Morningstar:1999rf}. Their masses depend on the two parameters of the theory, $\Lambda$ and $N$. 
Also from lattice calculations~\cite{Lucini:2010nv, Lucini:2001ej}, the lightest glueball masses approach  a constant at large $N$, and can be parametrized as
$m = (\alpha + \beta/N^2)\Lambda$ where $\alpha, \beta$ are order one parameters.
In general, we could also introduce the $\theta$-term in the above Lagrangian, which is $C$ even and $P$ odd.
It can mix the $0^{++}$ and $0^{-+}$ states and lightest glueball state is 
then not an eigenstate under $P$. 

We argue that within this simple setup the lightest hidden glueball state $\phi$ could be a candidate for dark matter~\cite{multicomponent}. 
It could be cosmologically long lived. As the lightest state, there is nothing in the hidden sector that $\phi$ could decay into.
It is possible for $\phi$ to decay into two gravitons, and this decay rate can be estimated as $\Gamma_{\phi} \sim m^5/M_{pl}^4 \sim \tau_U^{-1} (m/10^7\,{\rm GeV})^5$, where $\tau_U=10^{17} \,{\rm sec}$ is the age of our universe. The lifetime of $\phi$ against gravitational decay can be long enough if its mass is less than $10^7\,{\rm GeV}$.
Moreover, the hidden glueball $\phi$ particles could have the correct relic density and be (non-)relativistic enough as will be elaborated in the next section.
So far, we have not written down any interactions between the hidden section and the visible sector, which by gauge invariance is only possible 
in the form of higher dimensional operators. 
We will explore the resulting experimental bounds in an example where the hidden GDM $\phi$ decays into photons.

\begin{figure}[t]
\centerline{\includegraphics[width=0.8\columnwidth]{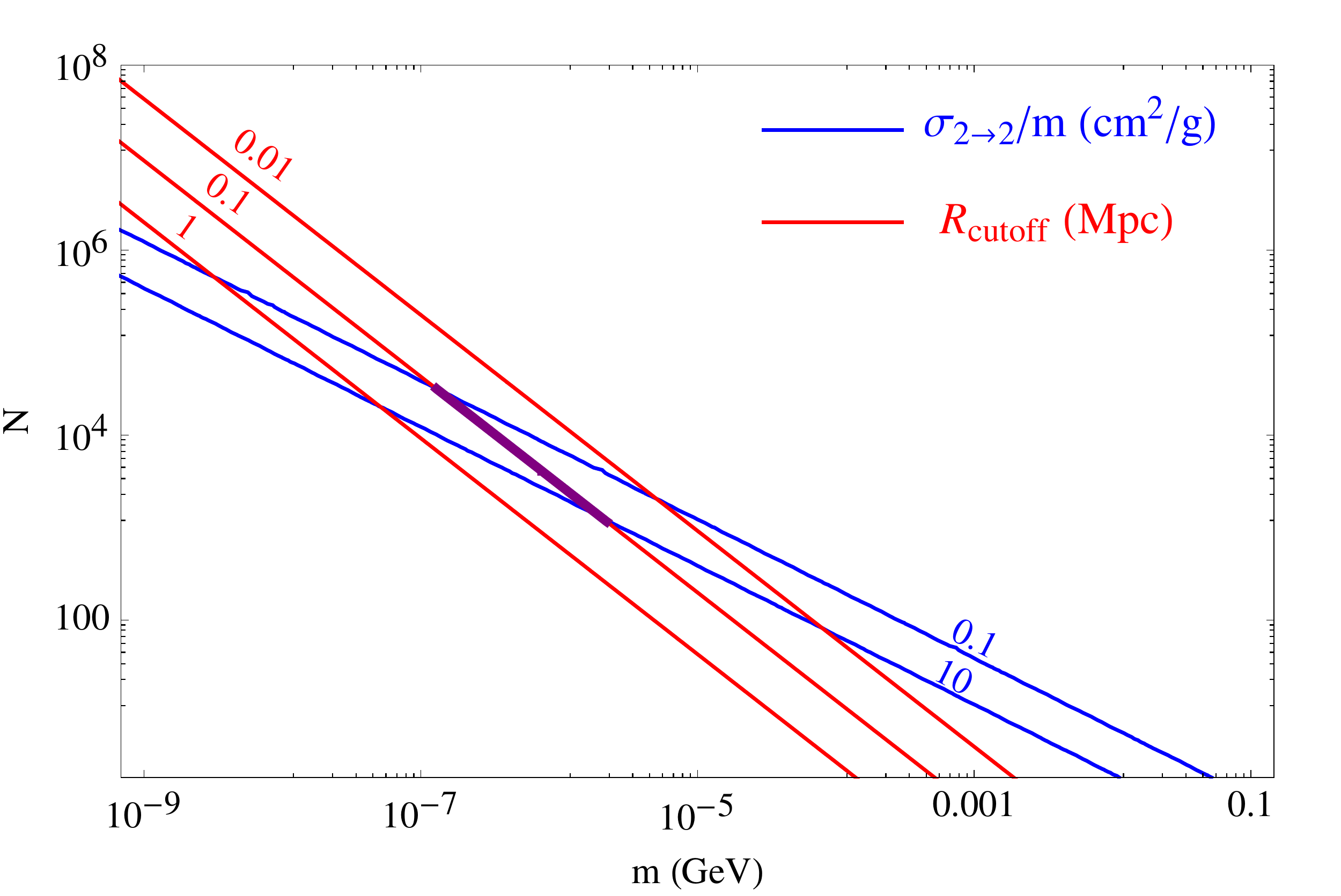}}
\caption{The parameter space of $m$ versus $N$ where the lightest hidden glueball could be a self-interacting and/or warm DMC.
The two blue curves correspond to constant values of DM self interaction cross section, $\sigma_{2\to2}/m = 0.1, 10\,{\rm cm^2/gram}$, respectively. 
Self-interacting DM lives between the blue curves.
The red curves correspond to constant values of damping scale in the power spectrum, $R_{\rm cutoff}=0.01, 0.1, 1\,$Mpc, respectively. 
Warm DM lives along the middle red curve.
The glueball dark matter can be both self-interacting and warm at the intersection of the two regions (thick purple curve).
}\label{sss}
\end{figure}

\smallskip
\noindent{\bfseries Self-interacting Dark Matter.}
The effective potential of a real scalar $\phi$ takes the form
\begin{eqnarray}\label{Veff}
V(\phi) = \frac{1}{2} m^2 \phi^2 + \frac{1}{3!}\lambda_3 \phi^3 + \frac{1}{4!}\lambda_4 \phi^4 + \frac{1}{5!}\lambda_5 \phi^5 + \cdots \ , \ \ 
\end{eqnarray}
where the $\cdots$ represent higher power terms. It is useful to consider the large $N$ behavior of these couplings,
\begin{eqnarray}
\lambda_3 = \frac{\kappa_3 m}{N}, \ \ \ \lambda_4 = \frac{\kappa_4}{N^2}, \ \ \ \lambda_5 = \frac{\kappa_5}{m N^3} \ ,
\end{eqnarray}
where $\kappa_{3,4,5}$ are order one parameters to be determined from non-perturbative calculations. 
From these interactions, we could obtain the $2\to2$ elastic scattering cross section of $\phi$ as a function of the two model parameters, 
$m (\Lambda)$ and $N$, $\sigma_{2\to2} \sim {1}/({m^2 N^4})$.
The self-interacting dark matter scenario has been proposed~\cite{Spergel:1999mh} to reconcile the core/cusp problem in dwarf galaxy observations and simulations.
For this scenario to work, the elastic scattering cross section of dark matter must lie in the range $0.1\,{\rm cm^2/gram} < \sigma_{2\to2}/m < 10\,{\rm cm^2/gram}$.
This requirement puts a correlated constraint on $m$ and $N$,
\begin{eqnarray}
m \sim 0.1 \,{\rm GeV} \cdot N^{-4/3} \ .
\end{eqnarray}
This region is shown between the blue curves in Fig.~\ref{sss}.

\medskip
\noindent{\bfseries Self Heating and Warm Dark Matter.}
In addition to elastic scattering, the effective interactions in (\ref{Veff}) also allow $\phi$ to have the inelastic $3\leftrightarrow2$ annihilation, which changes the $\phi$ particle number. The analog of cross section could be estimated as
$\sigma_{3\to2} \sim {1}/({m^5 N^6})$.
The $3\to2$ reaction rate is given by $\Gamma_{3\to2}= n_\phi^2 \sigma_{3\to2}$, where $n_\phi$ is the $\phi$ number density in the universe.
This interaction could play an important role on the velocity dispersion of dark matter in the early universe, because after each $3\to2$ reaction the two outgoing $\phi$ particles are relativistic. If this process has a larger reaction rate than the Hubble expansion, the annihilation will keep heating up the $\phi$ particles until it reaches the balance with the inverse process where two energetic $\phi$'s annihilate into three. 
In this model, there are no interactions for $\phi$ and SM particles to exchange heat in equilibrium~\cite{unique}, the entropy of the $\phi$ particles is conserved, $\frac{d}{da}[(\rho_\phi+p\phi) a^3/T]=0$. For non-relativistic $\phi$'s, {\it i.e.}, $T_\phi\ll m$, one could derive 
\begin{equation}\label{Ta}
T_\phi(a) \simeq T_\phi(a_0) \left(1+ \frac{3T_\phi(a_0)}{m}\ln\frac{a}{a_0} \right)^{-1}\ ,
\end{equation}
where $a$ is the Hubble radius at given time in the early universe ($a=1$ today), and $a_0<a$ corresponds to an earlier time. 
This means the $\phi$ particles thermalize to a temperature which drops more slowly than $1/(\ln a)$ with the expansion of the universe, as first noted in~\cite{Carlson:1992fn}.
In contrast, the temperature of the photons falls as $T_\gamma \sim 1/a$. 
leading to the interesting possibility that the hidden and SM sectors have their own temperatures and evolve separately.

\begin{figure}[t]
\centerline{\includegraphics[width=0.8\columnwidth]{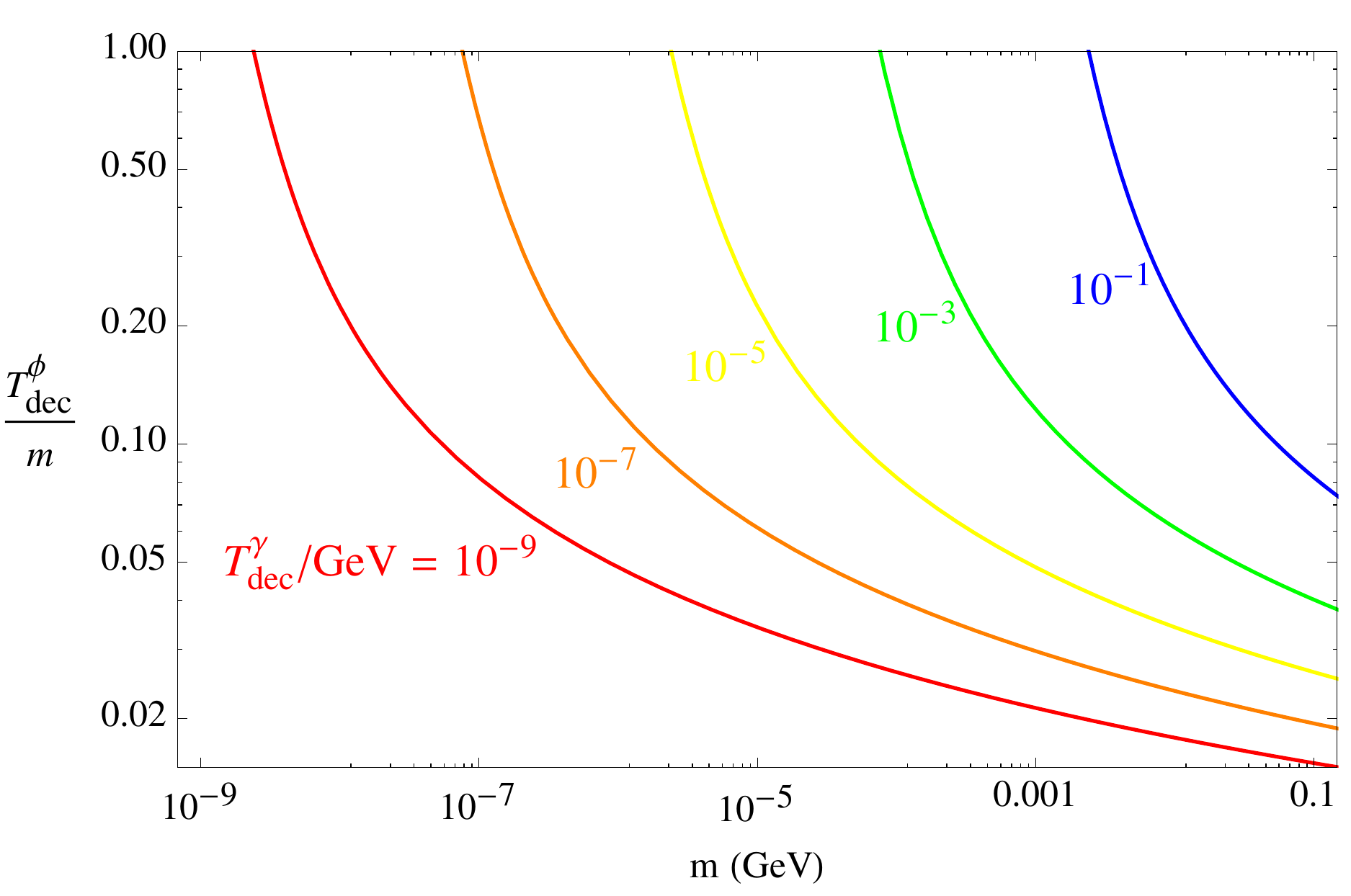}}
\caption{Ratio of temperature $T_\phi$ to the mass $m$ of $\phi$ particles at the decoupling of $3\to2$ annihilation that could give the correct dark matter relic density.
The curves correspond to different photon temperatures ($T_{dec}^\gamma$) at this epoch.
Roughly, $T_\phi$ is only one order of magnitude below the mass, and the $\phi$ particles remain heated before the decoupling.}\label{Tdec}
\end{figure}

It is useful to expand the energy density and pressure of $\phi$ to next order in $T_\phi/m$, $\rho_\phi = m n_\phi \left(1+3T_\phi/(2m) \right)$, $p_\phi = m n_\phi T_\phi/m$.
With this one can obtain the evolution equation of $n_\phi$ as a function of $a$,
\begin{eqnarray}
\frac{d(n_\phi a^3)}{da} \simeq - \frac{(n_\phi a^3)}{a} \frac{3T_\phi}{m}  \ .
\end{eqnarray}
The message here is that the number density of $\phi$ dilutes faster than $a^{-3}$, thus the total number of $\phi$ is still 
decreasing while the $3\to2$ annihilation is in equilibrium. The consumption of $\phi$'s is used to 
maintain the temperature of the remaining $\phi$ particles.
The final
DM relic density is given by $n_\phi$ at the decoupling of $3\to2$ annihilation. 
In Fig.~\ref{Tdec}, we show the ratio of the decoupling temperature $T^\phi_{dec}$ to the mass of $\phi$ that is needed to give the correct dark matter relic density, for different values of the photon temperature at this epoch~\cite{initialcondition}.

Before the $3\to2$ decoupling, the temperature $T_\phi$ stays roughly one order of magnitude below the mass $m$. The strongly coupled $\phi$ particles form a fluid with a large speed of sound $c_s = \sqrt{2T_\phi/(3m)}\sim 0.3 c$. It allows the perturbations to 
the  density of $\phi$ within one Hubble patch to be smoothed out efficiently via collisional damping, thus 
offering the opportunity for $\phi$ to be a warm
DMC.

To find when the $3\to2$ process decouples, or the corresponding temperature of photon $T^\gamma_{dec}$, we
first express $3\to2$ rate in terms of the photon temperature,
$\Gamma_{3\to2} = n_\phi^2 \sigma_{3\to2} \simeq {10^{-17} {\rm GeV}^2\, T_\gamma^6}/({m^7 N^6})$.
When it is equal to the Hubble rate, we get the photon temperature at $3\to2$ decoupling
$T_{dec}^\gamma \simeq 1\,{\rm keV} \left[ {m}/({1\,{\rm keV}}) \right]^{7/4} \left[ {N}/({10^4}) \right]^{3/2}$.
The collisional damping length scale (measured today) is determined by the Hubble radius at the $3\to2$ decoupling 
\begin{eqnarray}\label{Rcd}
R_{cd} = \frac{1}{H \left(T_{dec}^\gamma \right)} \frac{T_{dec}^\gamma}{2.7\,\rm K} \simeq 0.1\,{\rm Mpc} \left( \frac{1\,{\rm keV}}{T_{dec}^\gamma} \right) \ .
\end{eqnarray}
After the $3\to2$ decoupling, the temperature of $\phi$ will drop as $1/a^2$ such that the velocity redshifts as $1/a$. We calculate the free streaming length of $\phi$ particles from this time, $t^{3\to2}_{dec}$, to the time of matter-radiation equality, $t_{eq}$. This corresponds to the collisionless damping scale,
\begin{eqnarray}
R_{fs} = \int_{t_{dec}}^{t_{eq}} \frac{v(t)}{a(t)} dt = \frac{2 v_{eq} t_{eq}}{a_{eq}} \ln\left[\frac{a_{eq}}{a_{dec}} \frac{1+\sqrt{1+v_{eq}^2}}{1+\sqrt{1+v_{dec}^2}}\right].\nonumber 
\end{eqnarray}
At matter-radiation equality $t_{eq}=2\times10^{12}$\,sec, $a_{eq} = 1/(1+z_{eq})$, $z_{eq} \simeq 3360$, and $v_{eq} = v_{dec} a_{dec}/a_{eq}$.
In principle, the distance $\phi$ travels would be even shorter than $R_{fs}$, because of the $2\to2$ scatterings which if frequent would make the $\phi$ particles diffuse rather than free stream. 
In practice, we find that for most of the parameter space of interest to this study, $R_{fs}\lesssim R_{cd}$. Therefore, it is $R_{cd}$ in (\ref{Rcd}) that determines the actual damping scale $R_{\rm cutoff}$ in the dark matter power spectrum. 

For $\phi$ to be the warm dark matter which solves the missing satellite problem, it is required 
that 
$R_{\rm cutoff} = R_{cd} \sim 0.1\,$Mpc~\cite{Viel:2013apy}. The contours of fixed $R_{\rm curoff}$ are shown by the red curves in Fig.~\ref{sss}. 
We further find that for $m \in (0.1, 10)$\,keV and $N \in (10^5, 10^3)$
(along the thick purple curve),
the hidden glueball $\phi$ dark matter qualifies to be both self interacting and warm, thus 
plausibly
solving all the small scale structure problems.

Moreover, if the dark matter still have non-negligible velocity and fast $2\to2$ self interactions during the formation of the cosmic microwave background (CMB), it might leave an imprint in the CMB spectrum. We leave this interesting possibility for a future detailed study.

\medskip
\noindent{\bfseries Compact Boson Stars.}
So far, we have not considered any interactions between the 
hidden $SU(N)$ sector and SM particles. In the absence of such interactions, we would look for the dark matter only through gravitational effects. 
It has been shown that the 
dark scalar field could have Bose-Einstein condensation and 
form massive compact objects such as boson stars~\cite{Colpi:1986ye, Eby:2015hsq}.
This may result in very dramatic gravitational
effects in our universe today such as microlensing~\cite{Paczynski:1985jf, Griest:1990vu}.

The mass range of the boson star depends on whether the self-interaction of $\phi$ is repulsive or attractive.
The size of the boson star is typically much larger than the inverse of the glueball mass.
In the hidden glueball model 
Eq.~(\ref{Veff}), at low momentum transfer the effective coupling of the $\phi^4$ self interaction is 
\begin{equation}
\lambda_{\rm eff} = \lambda^2_3/(2m^2) + \lambda_4 = (\kappa_3^2/2 + \kappa_4)/N^2 \ .
\end{equation}
Non-perturbative calculations are needed to 
reliably determine the size and signs of $\kappa_3, \kappa_4$, and in turn the fate of the condensate.

The opportunity to observe the microlensing effect
arises if there  is  repulsive self interactions for the $\phi$ field, with $\lambda_{\rm eff}>0$.
In this case, it has been calculated~\cite{Eby:2015hsq} that the boson star mass from condensation lies in the range 
$1-10^{8} M_{\odot}$, for the glueball dark matter with mass from GeV to 0.1 keV scale.
In particular, in the interesting window of Fig.~\ref{sss} where the dark matter is both self-interacting and warm,
the corresponding boson star mass is between $10^{6}-10^8 M_{\odot}$.
On the other hand, if $\lambda_{\rm eff}<0$, the boson star mass would be too small to have an observable effect.

\medskip
\noindent{\bfseries Interactions with the SM Through Higher Dimensional Operators.}
In general, there may exist interactions between the hidden sector and the SM sector.  This may allow the glueball dark matter to be discovered through means
other than gravitational effects. 
However, we do not want to introduce other particles just to facilitate these interactions, since as explained before, we want to explore
how far our set up with just a simple pure $SU(N)$ gauge theory can go in addressing the DM issue.
So, without introducing additional particles, gauge invariance dictates that these interactions may arise via higher dimensional operators, 
$\mathcal{L}_{int} =(1/M^n){H_{\mu\nu} H^{\mu\nu} \mathcal{O}_{SM}}$,
where $M$ is the cutoff scale. 
There are many choices for the $\mathcal{O}_{SM}$ part. Here we discuss one representative which couples the hidden sector directly to photons
\begin{equation}\label{int}
\mathcal{L}_{int} =\frac{1}{M^4} H_{\mu\nu} H^{\mu\nu} (F_{\alpha\beta} F^{\alpha \beta}) \to \frac{N m^3}{M^4} \phi F_{\alpha\beta} F^{\alpha \beta} \ ,
\end{equation}
where $F$ is the photon field strength. In the second step, we go to the low scale 
where $\phi$ is the lightest glueball field. 
In the following, we choose the value of $N$ making $\phi$ a self-interacting dark matter, 
$N\simeq{\rm Max}\left[(m/0.1\,{\rm GeV})^{-3/4}, 2\right]$.
It is also worth noting that the effective interaction of $\phi$ is proportional to powers of its mass $m^3$.

\begin{figure}[t]
\centerline{\includegraphics[width=0.8\columnwidth]{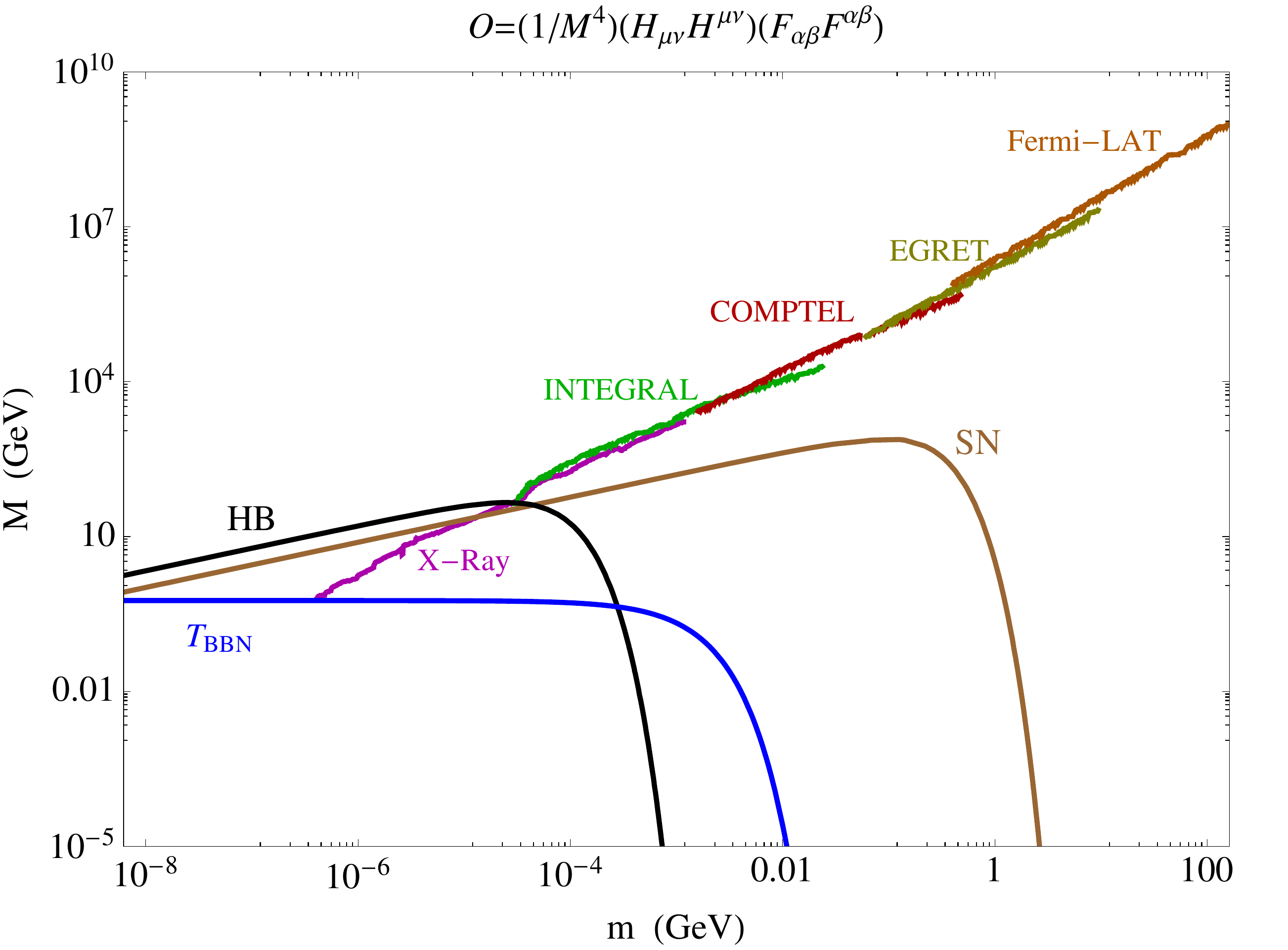}}
\caption{Lower bounds on the cutoff scale $M$.
Cosmic ray photon observations constrains glueball dark matter decay into photons, and from right to left, the curves correspond to constraints from Fermi-LAT, EGRET, COMPTEL, INTEGRAL, X-ray.
The black (brown) solid curve is the lower bound on $M$ from the energy loss argument of HB (SN).
The blue curve represents the requirement that the hidden sector is not thermalized with the SM sector below the BBN temperature.
}\label{bounds}
\end{figure}

From Eq.~(\ref{int}), the decay rate of $\phi$ into two photons is, 
\begin{equation}
\Gamma_{\phi\to\gamma\gamma} = \frac{N^2 m^9}{4\pi M^8} \ .
\end{equation}
There are experimental searches for monochromatic photon from decaying dark matter, from cosmic gamma rays to X rays and even extragalactic background lights~\cite{Ackermann:2015lka, Strong:2004de, Bouchet:2008rp, Essig:2013goa, Arias:2012az}. They give the strongest constraints on the scale $M$ for the dark matter $\phi$ mass 
above $\sim$\,100\,keV. 
We show these constraints in Fig.~\ref{bounds}.

For lower $\phi$ masses, we find the energy loss constraints of stars place a stronger lower limit. The relevant 
reaction is the Primakoff type process $e +\gamma \to e + \phi$.
The cross section was calculated in~\cite{Dicus:1978fp},
\begin{equation}\label{primakoff}
\sigma v = 64 \pi \alpha \frac{\omega \Gamma_{\phi\to\gamma\gamma}}{m^2} \frac{(\omega^2 - m^2)^{1/2} (\omega - m)}{(m^2 - 2 \omega m)^2} \ ,
\end{equation}
where $\omega$ is the energy of the incoming photon and $m$ is the mass of glueball dark matter.
To calculate the rate of energy loss from the star via $\phi$ emission, we first average the $\sigma v \cdot \omega$ over the thermal photon energy distribution, and then the energy loss rate per unit volume is given by $\Phi = n_e n_\gamma \langle \sigma v \cdot \omega \rangle$. 
We consider the energy loss argument~\cite{Pospelov:2008jk} of horizontal branch stars (HB) and the cooling of type-II supernova (SN).
For HB, the core temperature is 10\,keV, the mass density is $10^4 {\rm gram/cm^3}$, and the energy loss rate per 
unit volume is required to be $\Phi<10^{-42}\,{\rm MeV}^5$.
For SN, the core temperature is 30\,MeV, both photon and electrons are thermalized, and the energy loss rate is required to be $\Phi<10^{-14}\,{\rm MeV}^5$.
Their constraints on $M$ (lower bound) is shown in Fig.~\ref{bounds}.
Not-too-much energy loss of HB sets the strongest lower bound on $M$ for $\phi$ mass below $\sim100$\,keV. 
For the model to be realistic in cosmology,  the hidden sector must not thermalize with the SM sector, at least not since the onset of BBN. We find this to be a subdominant constraint (shown by the blue curve in Fig.~\ref{bounds}).

The operators in Eq.~(\ref{int}) not only lead the glueball dark matter particle to decay, but also allows it to scatter with SM particles by virtue of the expansion
$H_{\mu\nu} H^{\mu\nu} \sim Nm^3 \phi + m^2 \phi^2 + \cdots$.
Given the above lower bounds on the cutoff scale $M$, we find the direct detection cross section for the glueball dark matter is more than tens of orders of magnitude below the 
current LUX bound~\cite{Akerib:2015rjg}. This is consistent with the null results so far in the direct detections. It also implies that if the future direct detection experiments discovers the dark matter, it cannot originate from our dark matter candidate~\cite{commentonDD}.

From Fig.~\ref{bounds}, we find that for the dark matter mass $m$ in the range keV to MeV, the cutoff $M$ is allowed to be as low as the weak (or TeV) scale.
The effective operator in Eq.~(\ref{int}) could be generated by integrating out a heavy particle $X$ in the ultraviolet theory, which carries both electromagnetic charge and color under the hidden $SU(N)$ gauge group. If a pair of $X\bar X$ can be produced at colliders, they would eventually form a heavy $X$-onium bound state and annihilate away into the hidden glueball dark matter or photons. The final states will exhibit exotic signatures like the quirks~\cite{Kang:2008ea, Nussinov:2014xua}.

Furthermore, if the heavy $X$ particle is a fermion and also carries color under the $SU(3)_c$ of QCD,
the effective Lagrangian will contain an operator $(1/M^4)(H \widetilde H) (G \widetilde G)$ (similar to Eq.~(12) of Ref.~\cite{Kuzmin:1992up}). In the presence of the $\theta H \widetilde H$ term from the hidden $SU(N)$ theory, it induces an effective $\theta_{\rm QCD} G \widetilde G$ term, with $\theta_{\rm QCD} \sim (m/M)^4 \theta$, and makes a contribution to the neutron electric dipole moment (nEDM).
The important point we want to make here is that nEDM bound does not require the $\theta$ parameter of $SU(N)$ to be unnaturally small, unlike $\theta_{\rm QCD}$.
The current experimental upper bound on nEDM of around $10^{-26}$\,e\,cm~\cite{Baker:2006ts} translates, by the arguments of~\cite{Crewther:1979pi}, into $\theta_{\rm QCD} \lesssim 10^{-13}$. From the above relation between $\theta_{\rm QCD}$ and $\theta$, we find that $\theta$ is allowed to be order one if $m/M\lesssim 10^{-3}$,
which is always satisfied from Fig.~\ref{bounds}.

\medskip
\noindent{\bfseries Summary.}
In this paper, we investigate the physics of $SU(N)$ glueball dark matter from a
pure gauge theory non-abelian hidden sector. 
In spite of the simple setup, with few parameters,  there are quite a few novel features of this dark matter candidate.
We have discussed the conditions for it to be self-interacting and/or warm dark matter.  
The glueball dark matter could also condense into more compact objects like boson stars and be observed by gravitational lensing effects.
Therefore, our model can naturally accommodate the fact that there is only gravitational evidence for dark matter so far~\cite{Bertone:2004pz, Jungman:1995df}.
It could also interact with the standard model sector via higher dimensional operators and subject to traditional
direct searches for light scalar dark particles. 
The direct detection cross section of the glueball dark matter is constrained to be well below the 
experimental sensitivity,
now as well as for the foreseeable future.
We also comment on the possible UV origin of the higher dimensional operators leading to interesting collider signatures.

\smallskip

\noindent{\it Acknowledgements.}
One of us (AS) wants to thank Alex Kusenko for useful conversations.
The work of AS is supported in part by the DOE Grant No. DE-AC-02-98CH10886. The work of YZ is supported by the Gordon and Betty Moore Foundation through Grant \#776 to the Caltech Moore Center for Theoretical Cosmology and Physics, and by the DOE Grant Nos.~DE-FG02-92ER40701 and DE-SC0010255.


\begin{thebibliography}{99}

\bibitem{otherworks}
We are aware of several other works which study dark matter with a non-abelian dark sector, however, these works involve in addition rather elaborate other intricacies 
with significantly different phenomenology from this study; see, {\it e.g}, ~\cite{Boddy:2014yra, Boddy:2014qxa, Buen-Abad:2015ova, Yamanaka:2014pva}.


\bibitem{Boddy:2014yra} 
  K.~K.~Boddy, J.~L.~Feng, M.~Kaplinghat and T.~M.~P.~Tait,
  Phys.\ Rev.\ D {\bf 89}, no. 11, 115017 (2014)
  doi:10.1103/PhysRevD.89.115017
  [arXiv:1402.3629 [hep-ph]].

\bibitem{Boddy:2014qxa} 
  K.~K.~Boddy, J.~L.~Feng, M.~Kaplinghat, Y.~Shadmi and T.~M.~P.~Tait,
  Phys.\ Rev.\ D {\bf 90}, no. 9, 095016 (2014)
  doi:10.1103/PhysRevD.90.095016
  [arXiv:1408.6532 [hep-ph]].

\bibitem{Buen-Abad:2015ova} 
  M.~A.~Buen-Abad, G.~Marques-Tavares and M.~Schmaltz,
  Phys.\ Rev.\ D {\bf 92}, no. 2, 023531 (2015)
  doi:10.1103/PhysRevD.92.023531
  [arXiv:1505.03542 [hep-ph]].


\bibitem{Yamanaka:2014pva} 
  N.~Yamanaka, S.~Fujibayashi, S.~Gongyo and H.~Iida,
  arXiv:1411.2172 [hep-ph].
  

\bibitem{Cornwall:1982zn} 
  J.~M.~Cornwall and A.~Soni,
  Phys.\ Lett.\ B {\bf 120}, 431 (1983).
  doi:10.1016/0370-2693(83)90481-1

\bibitem{Morningstar:1999rf} 
  C.~J.~Morningstar and M.~J.~Peardon,
  Phys.\ Rev.\ D {\bf 60}, 034509 (1999)
  doi:10.1103/PhysRevD.60.034509
  [hep-lat/9901004].

\bibitem{Lucini:2010nv} 
  B.~Lucini, A.~Rago and E.~Rinaldi,
  JHEP {\bf 1008}, 119 (2010)
  doi:10.1007/JHEP08(2010)119
  [arXiv:1007.3879 [hep-lat]].

\bibitem{Lucini:2001ej} 
  B.~Lucini and M.~Teper,
  JHEP {\bf 0106}, 050 (2001)
  doi:10.1088/1126-6708/2001/06/050
  [hep-lat/0103027].

\bibitem{multicomponent}
If after mixing of the $0^{++}$ and $0^{-+}$ glueball states, the heavier mass eigenstate is kinematically forbidden to decay into two $\phi$'s, it can also be stable and be the dark matter. In this case, we could have two components of dark matter existing in nature.

\bibitem{Spergel:1999mh} 
  D.~N.~Spergel and P.~J.~Steinhardt,
  Phys.\ Rev.\ Lett.\  {\bf 84}, 3760 (2000)
  doi:10.1103/PhysRevLett.84.3760
  [astro-ph/9909386].

\bibitem{unique}
Gauge invariance dictates the interactions between the SM and hidden sector to take the form $H_{\mu\nu} H^{\mu\nu} \mathcal{O}_{SM}$. They will cause the dark matter $\phi$ to decay thus are highly constrained as we show below. This makes the early universe history of dark matter in our model very different from the one considered in~\cite{Hochberg:2014kqa}.

\bibitem{Hochberg:2014kqa} 
  Y.~Hochberg, E.~Kuflik, H.~Murayama, T.~Volansky and J.~G.~Wacker,
  Phys.\ Rev.\ Lett.\  {\bf 115}, no. 2, 021301 (2015)
  doi:10.1103/PhysRevLett.115.021301
  [arXiv:1411.3727 [hep-ph]].


\bibitem{Carlson:1992fn} 
  E.~D.~Carlson, M.~E.~Machacek and L.~J.~Hall,
  Astrophys.\ J.\  {\bf 398}, 43 (1992).
  doi:10.1086/171833

\bibitem{initialcondition}
The initial conditions that give the desired values of $T^\phi_{dec}$ and $T^\gamma_{dec}$ might be set by reheating the SM and dark sectors to different temperatures after the inflation. See, {\it e.g.}, 
  A.~E.~Faraggi and M.~Pospelov,
  Astropart.\ Phys.\  {\bf 16}, 451 (2002)
  doi:10.1016/S0927-6505(01)00121-9
  [hep-ph/0008223].

  
\bibitem{Viel:2013apy} 
  M.~Viel, G.~D.~Becker, J.~S.~Bolton and M.~G.~Haehnelt,
  Phys.\ Rev.\ D {\bf 88}, 043502 (2013)
  doi:10.1103/PhysRevD.88.043502
  [arXiv:1306.2314 [astro-ph.CO]].
  
\bibitem{Colpi:1986ye} 
  M.~Colpi, S.~L.~Shapiro and I.~Wasserman,
  Phys.\ Rev.\ Lett.\  {\bf 57}, 2485 (1986).
  doi:10.1103/PhysRevLett.57.2485
  
\bibitem{Eby:2015hsq} 
  J.~Eby, C.~Kouvaris, N.~G.~Nielsen and L.~C.~R.~Wijewardhana,
  arXiv:1511.04474 [hep-ph].


\bibitem{Paczynski:1985jf} 
  B.~Paczynski,
  Astrophys.\ J.\  {\bf 304}, 1 (1986).
  doi:10.1086/164140


\bibitem{Griest:1990vu} 
  K.~Griest,
  Astrophys.\ J.\  {\bf 366}, 412 (1991).
  doi:10.1086/169575


\bibitem{Ackermann:2015lka} 
  M.~Ackermann {\it et al.} [Fermi-LAT Collaboration],
  Phys.\ Rev.\ D {\bf 91}, no. 12, 122002 (2015)
  doi:10.1103/PhysRevD.91.122002
  [arXiv:1506.00013 [astro-ph.HE]].
\bibitem{Strong:2004de} 
  A.~W.~Strong, I.~V.~Moskalenko and O.~Reimer,
  Astrophys.\ J.\  {\bf 613}, 962 (2004)
  doi:10.1086/423193
  [astro-ph/0406254].
\bibitem{Bouchet:2008rp} 
  L.~Bouchet, E.~Jourdain, J.~P.~Roques, A.~Strong, R.~Diehl, F.~Lebrun and R.~Terrier,
  Astrophys.\ J.\  {\bf 679}, 1315 (2008)
  doi:10.1086/529489
  [arXiv:0801.2086 [astro-ph]].
\bibitem{Essig:2013goa} 
  R.~Essig, E.~Kuflik, S.~D.~McDermott, T.~Volansky and K.~M.~Zurek,
  JHEP {\bf 1311}, 193 (2013)
  doi:10.1007/JHEP11(2013)193
  [arXiv:1309.4091 [hep-ph]].
\bibitem{Arias:2012az} 
  P.~Arias, D.~Cadamuro, M.~Goodsell, J.~Jaeckel, J.~Redondo and A.~Ringwald,
  JCAP {\bf 1206}, 013 (2012)
  doi:10.1088/1475-7516/2012/06/013
  [arXiv:1201.5902 [hep-ph]].
  
  
  

\bibitem{Dicus:1978fp} 
  D.~A.~Dicus, E.~W.~Kolb, V.~L.~Teplitz and R.~V.~Wagoner,
  Phys.\ Rev.\ D {\bf 18}, 1829 (1978).
  doi:10.1103/PhysRevD.18.1829

\bibitem{Pospelov:2008jk} 
  M.~Pospelov, A.~Ritz and M.~B.~Voloshin,
  Phys.\ Rev.\ D {\bf 78}, 115012 (2008)
  doi:10.1103/PhysRevD.78.115012
  [arXiv:0807.3279 [hep-ph]].

\bibitem{Akerib:2015rjg} 
  D.~S.~Akerib {\it et al.} [LUX Collaboration],
  arXiv:1512.03506 [astro-ph.CO].

\bibitem{commentonDD}
A positive direct detection signal would imply our hidden glueball dark matter cannot comprise all the dark matter relic density. There might be other components of dark matter.

\bibitem{Kang:2008ea} 
  J.~Kang and M.~A.~Luty,
  JHEP {\bf 0911}, 065 (2009)
  doi:10.1088/1126-6708/2009/11/065
  [arXiv:0805.4642 [hep-ph]].
 
\bibitem{Nussinov:2014xua} 
  S.~Nussinov,
  arXiv:1410.6409 [hep-ph].
 

\bibitem{Kuzmin:1992up} 
  V.~A.~Kuzmin, M.~E.~Shaposhnikov and I.~I.~Tkachev,
  Phys.\ Rev.\ D {\bf 45}, 466 (1992).
  doi:10.1103/PhysRevD.45.466
 
\bibitem{Baker:2006ts} 
  C.~A.~Baker {\it et al.},
  Phys.\ Rev.\ Lett.\  {\bf 97}, 131801 (2006)
  doi:10.1103/PhysRevLett.97.131801
  [hep-ex/0602020].



\bibitem{Crewther:1979pi} 
  R.~J.~Crewther, P.~Di Vecchia, G.~Veneziano and E.~Witten,
  Phys.\ Lett.\ B {\bf 88}, 123 (1979)
  [Phys.\ Lett.\ B {\bf 91}, 487 (1980)].
  doi:10.1016/0370-2693(79)90128-X

\bibitem{Bertone:2004pz} 
  G.~Bertone, D.~Hooper and J.~Silk,
  Phys.\ Rept.\  {\bf 405}, 279 (2005)
  doi:10.1016/j.physrep.2004.08.031
  [hep-ph/0404175].

\bibitem{Jungman:1995df} 
  G.~Jungman, M.~Kamionkowski and K.~Griest,
  Phys.\ Rept.\  {\bf 267}, 195 (1996)
  doi:10.1016/0370-1573(95)00058-5
  [hep-ph/9506380].
  
\end{thebibliography}
\end{document}